  \def   \ni {\noindent}

  \def   \ssk {\vskip  5truept}
  \def   \sk  {\vskip 10truept}
  \def   \bsk {\vskip 15truept}

  \def   \newline {\hfil\break}

\input psfig.tex
  \magnification=1000
  \hsize 5truein
  \vsize 8truein
  \font\abstract=cmr8
  
  \font\text=cmr10     
  \font\affiliation=cmssi10
  \font\author=cmss10
  
  \font\title=cmssbx10 scaled\magstep2

  \def\ref{\par\noindent\hangindent 15pt}
  \nopagenumbers
  \null
  \vskip 3.0truecm
  \baselineskip = 12pt
  
  {\title                       
  \ni Radio and Far-Infrared Extragalactic Sources at Planck Frequencies
  }
  \bsk \bsk
  {\author                              
  \ni GIANFRANCO DE ZOTTI$^1$ AND LUIGI TOFFOLATTI$^{1,2}$
  }
\sk 
\ni
 {\affiliation {}$^1$Osservatorio Astronomico di Padova, Italy 
  }

\ni
{\affiliation {}$^2$ Dep. de F\'\i{sica}, Universidad de Oviedo, Spain}
  \bsk
  \bsk
\def\lsim{\, \lower2truept\hbox{${< \atop\hbox{\raise4truept\hbox{$\sim$}}}$}\,}
\def\gsim{\, \lower2truept\hbox{${> \atop\hbox{\raise4truept\hbox{$\sim$}}}$}\,}
  \baselineskip = 9pt
  {\abstract                                   
\ni
We discuss the main uncertainties affecting estimates of small scale 
fluctuations due to extragalactic sources in the Planck Surveyor frequency 
bands. Conservative estimates allow us to confidently conclude that, in the 
frequency range 100--200 GHz, the contaminating effect of extragalactic 
sources is well below the expected anisotropy level of the cosmic microwave 
background (CMB), down to angular scales of at least $\simeq 
10'$. Hence, an accurate subtraction of foreground fluctuations 
is not critical for the determination of the CMB power spectrum up 
to multipoles $ \ell \simeq 1000$. In any case, Planck's wide frequency 
coverage will allow to carefully control foreground contributions.
On the other hand, the all sky surveys at 9 frequencies, spanning the 
range 30--900 GHz, 
will be unique  in providing complete samples comprising from several 
hundreds to many thousands of extragalactic sources, selected in an 
essentially unexplored frequency interval. New classes of sources may be 
revealed in these data. Extremely compact radio sources, whose radio emission 
is relativistically boosted both in intensity and in frequency will be 
particularly prominent. Within this frequency region, very 
compact synchrotron components become optically thin; the corresponding 
break frequency is a key parameter in models of the energy distribution.   
Crucial information will be provided to understand the 
nature of radio sources with strongly inverted spectra. Scenarios for 
the cosmological evolution of galaxies will be tested.
  }                                                  
   \sk
  \baselineskip = 12pt
  {\text                                         
  \ni 1. INTRODUCTION

\ni
Astrophysical foregrounds constitute an unavoidable fundamental limitation to 
measurements of primordial anisotropies of the Cosmic Microwave Background 
(CMB).  Due to the large beam size, the COBE/DMR data are, to some extent, 
contaminated by galactic emission (Kogut et al. 1996), whereas they are 
basically unaffected by extragalactic sources (Kogut et al. 1994; Banday et 
al. 1996). On the other hand, confusion noise due to discrete sources is 
likely to be the main limiting factor for experiments, like Planck, aimed at 
accurately determining the CMB power spectrum down to small angular scales 
(Tegmark \& Efstathiou 1996; Toffolatti et al. 1998).

In \S$\,2$ we review the main results of recent analyses of 
fluctuations due to extragalactic sources and discuss their uncertainties. 
In \S$\,3$ we comment on the astrophysical 
information on such sources that will be provided by Planck surveys. 
In \S$\,4$ we summarize our main conclusions. 
}
\bsk

  {\text 
  \ni 2. FLUCTUATIONS DUE TO DISCRETE EXTRAGALACTIC SOURCES

\ni
The contributions of discrete extragalactic sources to small scale 
fluctuations have been extensively discussed in the literature
(Franceschini et al. 1989; Tegmark \& Efstathiou 1996; Gawiser \& Smoot 1997; 
Toffolatti et al. 1998; Guiderdoni et al. 1998; Blain et al. 1998). 
The main results can be summarized as follows.  

The sharp rise 
of dust emission spectra with increasing frequency (typically 
$S_\nu \propto \nu^{3.5}$) determines a drastic change in the composition 
of the population of bright extragalactic sources above and 
below $\simeq 200\,$GHz: radio sources (mostly 
``flat''-spectrum radiogalaxies, quasars, BL Lacs, blazars) dominate at lower 
frequencies, and evolving dusty galaxies at higher frequencies.

\smallskip

\ni {\bf 2.1 Fluctuations due to radio sources}

\ni 
Estimates of fluctuations due to radio sources are based 
on extrapolations of evolutionary models which fit the observed counts 
at several frequencies from 408 MHz to 8.44 GHz. As such counts reach 
much deeper flux levels than achievable by Planck, the critical 
ingredient for the extrapolation is not the evolution model but rather 
the spectral behaviour at high frequencies.

In carrying out extrapolations in frequency, several effects need to be 
taken into account. On one side, 
the majority of sources with flat or inverted spectra at 5 GHz 
have spectral turnovers below 90 GHz (Kellermann \& Pauliny-Toth 1971; 
Owen \& Mufson 1977). This is not surprising since astrophysical 
processes work to steepen the high frequency source spectra. For a 
power law energy distribution of relativistic electrons 
[$N(E)\propto E^{-p}$, $p\simeq 2.5$] 
the synchrotron self-absorption coefficient is proportional to $\nu^{-(p+4)/2}$ 
(Rybicki \& Lightman 1979). Thus even the most 
compact synchrotron components become optically thin at high enough frequencies 
and the emission takes on a spectral index $\alpha \simeq (p-1)/2$ 
($S_\nu \propto \nu^{-\alpha}$).
A further steepening of the spectrum is produced by electron energy losses.

On the other side, 
high frequency surveys preferentially select sources with 
harder spectra. For power law differential source counts, $n(S,\nu_0) = 
k_0\,S^{-\gamma}$, and a Gaussian spectral index distribution with 
mean $<\alpha>_0$ and dispersion $\sigma$, the counts at a frequency $\nu$ 
are given by $n(S,\nu) = n(S,\nu_0)(\nu/\nu_0)^{\alpha_{\rm eff}}$ 
with (Kellermann 1964; Condon 1984):
$\alpha_{\rm eff}= <\alpha>_0 + \ln(\nu/\nu_0)\sigma^2(1-\gamma)^2/2$. 
Estimates neglecting the dispersion of spectral indices underestimate 
the counts by a factor $\exp[\ln^2(\nu/\nu_0)\sigma^2(1-\gamma)^2/2]$.
The spectral index distribution between 5 and 90 GHz determined by Holdaway et 
al. (1994) has $\sigma = 0.34$; for Euclidean counts, $\gamma =2.5$, 
the correction then amounts to about a factor of 3.

A good fraction of the observed spread of spectral indices is due to 
variability whose  rms amplitude, in the case of blazars,  
increases with frequency, reaching a factor of about 1.5 at a few hundred GHz
(Impey \& Neugebauer 1988). 
In some cases, variations by a factor of 2 to 3 have been observed at these 
frequencies (e.g. 3C345: Stevens et al. 1996; PKS$\,0528+134$: Zhang et al. 
1994). The highest frequency outbursts are expected to be associated to the 
earliest phases of the flare evolution. Since the rise of the flare is often 
rather abrupt (timescale of weeks), they were probably frequently missed.  

If we adopt a lognormal model for the distribution of variable fluxes, such 
that the probability $p(S)$ that a source of average flux $\overline{S}$ is 
observed to have a flux $S$ is:
$$p(S)\,dS= {1\over \sqrt{2\pi} \sigma_v} \exp{\left[-{(\ln S - 
\ln\overline{S})^2 \over 2\sigma_v^2}\right]} {dS \over S},$$
and if the differential source counts in the absence of variability are 
described by $n(\overline{S}) = k\,\overline{S}^{-\gamma}$, we get:
$$n(S) = k S^{-\gamma} \exp\left[{1\over 2} (\gamma -1)^2 \sigma_v^2\right]$$
For $\sigma_v = \ln 1.5$ and $\gamma =2.5$, counts are enhanced by a factor 
$\simeq 1.2$. This enhancement factor is likely a lower limit, since, as 
mentioned above, high frequency outbursts are likely to have been undersampled. 

An accurate modelling of all the above effects is impossible in the present 
data situation. However, the simple recipe adopted by Danese et al. (1987) and 
taken up by Toffolatti et al. (1998) (spectral index 
$\alpha =0$  for ``flat''-spectrum sources up 
to 200 GHz, and a subsequent steepening to $\alpha =0.75$) turns out to 
be rather successful. In fact, 
it allowed to reproduce, without any adjustment of the parameters, 
the deep counts at 8.44 GHz (Windhorst et al. 1993; Partridge et al. 1997), 
which were produced several years after the model. It is also consistent 
with the estimates of 90 GHz counts obtained by Holdaway et al. (1994) 
from sensitive 90 GHz observations of a large sample of 
``flat''-spectrum sources selected at 5 GHz. Actually, the numbers of 
sources brighter than 0.1 Jy and 1 Jy estimated by Holdaway et al. (1994) 
are almost a factor of 2 below the predictions by Toffolatti et al. (1998). 
However, the former results are really lower limits since they may 
underestimate the number of sources with strongly inverted spectra 
which are under-represented at 5 GHz.  

On the other hand, the counts predicted by Toffolatti et al. (1998) at  
frequencies above 100 GHz may be somewhat depressed by the assumption 
of a spectral break at 200 GHz for all ``flat''-spectrum sources 
while examples are known of sources keeping a flat or 
inverted spectrum up to 1000 GHz. The results presented 
in Figs. 1 and 2 are obtained adopting the latter value for the break 
frequency. At $\nu \gsim 300\,$GHz, 
however, counts are probably dominated by dusty galaxies.

\smallskip
\ni {\bf 2.2 Fluctuations due to dusty galaxies}

\ni 
Although the situation is rapidly improving, thanks to the 
deep ISO counts at $175\,\mu$m (Kawara et al. 1997; Puget et al. 1998), to  
the preliminary counts at $850\,\mu$m with SCUBA on JCMT (Smail et 
al. 1997; Hughes et al. 1998; Barger et al. 1998; Eales et al. 1998) 
and to the important constraints from measurements of the far-IR to 
mm extragalactic background (Schlegel et al. 1998; Hauser et al. 1998; 
Fixsen et al. 1998), 
first detected by Puget et al. (1996), current estimates are affected by 
bigger uncertainties than in the case of radio sources.

In fact, predicted counts have a higher responsiveness to the poorly known 
evolutionary properties, because of the 
boosting effect of the strongly negative K corrections.  The most extensive 
surveys, carried out by IRAS at $60\,\mu$m, cover a limited range in flux 
and are rather uncertain at the faint end (Hacking \& Houck 1987; 
Gregorich et al. 1995; Bertin et al. 1997). It is then not surprising that 
predictions of recent models differ by substantial factors (cf. Table 1).

\topinsert  
  \centerline {TABLE 1: Predicted galaxy counts in PLANCK HFI bands} 
\centerline {$\log N(>100\,\hbox{mJy})\ \hbox{sources}\,\hbox{sr}^{-1}$  }
  \ssk
$$\vbox{\hsize 2.5truein \settabs 5 \columns
  \hrule
\vskip5pt
\+ & $350\,\mu$m\qquad & $550\,\mu$m\qquad & $850\,\mu$m\qquad  & 
$1380\,\mu$m\qquad  \cr
\vskip5pt
\hrule
\vskip5pt
\+ Guiderdoni et al. (1998) &&&& \cr
\+ Q & 3.97 & 3.06 & 2.15 & 1.18 \cr
\+ A & 3.93 & 2.73 & 1.67 & 0.71 \cr
\+ B & 4.41 & 3.48 & 2.23 & 0.83 \cr
\+ C & 4.95 & 4.88 & 4.39 & 3.15 \cr
\+ D & 5.04 & 4.34 & 3.33 & 1.77 \cr
\+ E & 5.08 & 4.57 & 3.83 & 2.39 \cr
\smallskip
\+ Franceschini et al. (1997) &&&& \cr
\+ Opaque & 4.80 & 4.04 & 2.85 & 1.71 \cr
\+ No-evol & 4.02 & 3.14 & 2.07 & 0.97 \cr
\smallskip
\+ Rowan-Robinson (1998), minimal evolution &&&& \cr
\+ & 3.2\phantom{0} & 2.2\phantom{0} &  & \cr 
\vskip5pt
\hrule
  }$$

\endinsert

Again substantial extrapolations in frequency are required, and have 
to deal with the poor knowledge of the spectrum of galaxies in the mm/sub-mm 
region; the $1.3\,$mm/$60\,\mu$m flux ratios of galaxies are observed 
to span about a factor of 10 (Chini et al. 1995; Franceschini \& 
Andreani 1995). A bivariate $60\,\mu$m/$1.3\,$mm luminosity distribution 
has been obtained by Franceschini et al. (1998). 

\smallskip

\ni {\bf 2.3 Power spectrum of foreground fluctuations at different frequencies}

\ni
As pointed out in the proposals submitted to ESA for the High Frequency 
and Low Frequency Instruments (HFI and LFI), 
due to their high sensitivity, the experimental accuracy is  
effectively limited by astrophysical foregrounds.
On angular scales $\gsim 30'$, foreground fluctuations are dominated by 
Galactic emissions. COBE data indicate that these reach a minimum between  
50 and $90\,$GHz (Kogut et al. 1996; Kogut 1996). The models by Toffolatti 
et al. (1998) imply that the minimum in the foreground  
fluctuation spectrum moves to higher frequencies with 
increasing Galactic latitude and decreasing angular scale; for 
$|b| > 50^\circ$ and $\theta \simeq 30'$ it occurs at about 
100 GHz.

\topinsert
\par\noindent
\centerline{\psfig{figure=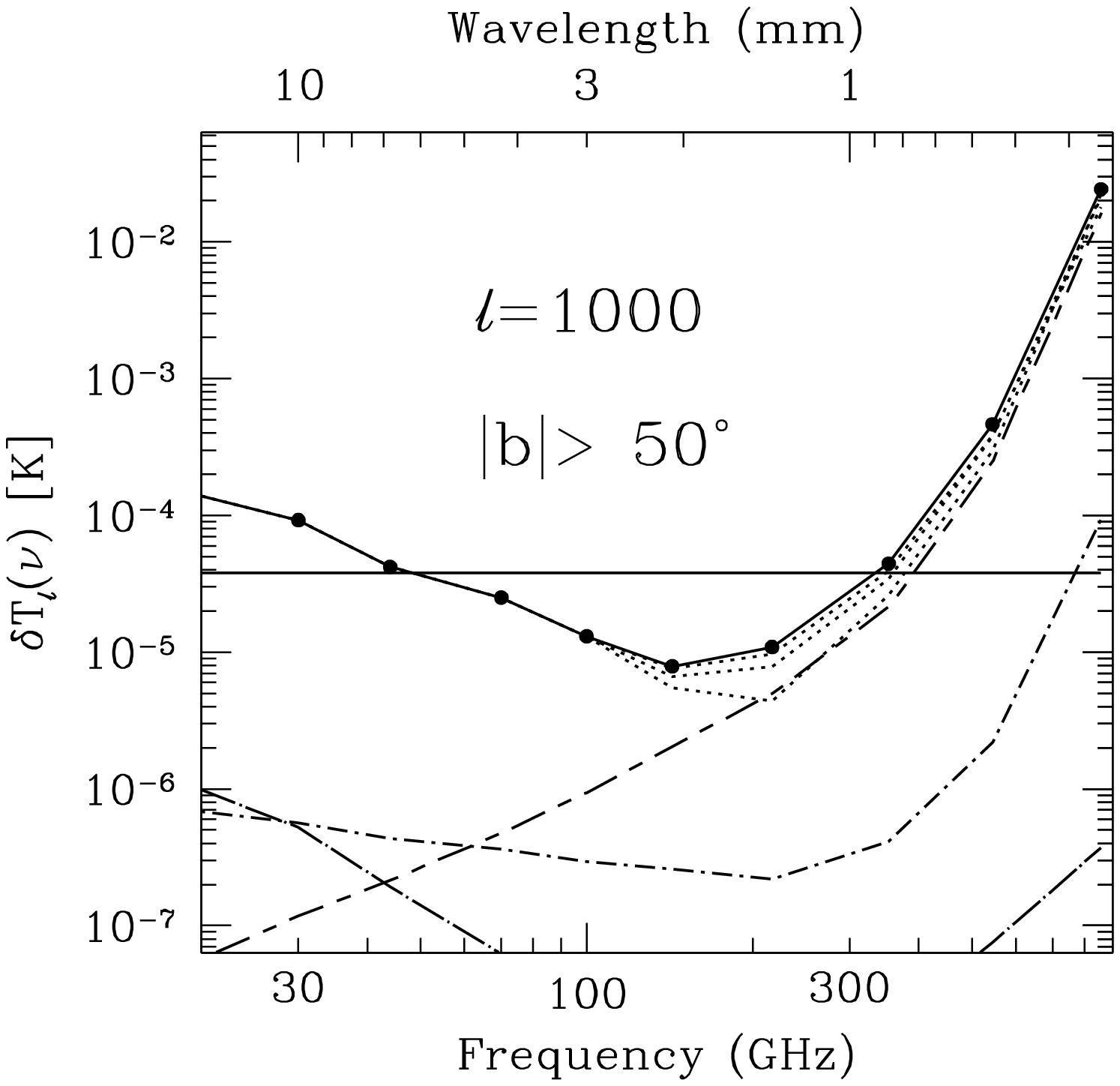,height=9.0truecm,width=12truecm}}

\leftskip=1truecm
\rightskip=1truecm
\vskip-2.truecm
\noindent
{\abstract FIGURE 1. 
Frequency dependence of CMB brightness fluctuations $\delta T_\ell = 
[\ell(\ell +1)C_\ell(\nu)/4\pi]^{1/2}$ for the multipole 
$\ell = 1000$ corresponding to an angular scale $\theta \simeq 180^\circ/\ell
\simeq 0.18^\circ \simeq 10'$. The dots$+$long dashes, dots$+$short dashes, 
and long$+$short dashes correspond to the mean contributions from Galactic 
foregrounds (synchrotron, free-free, and interstellar dust, respectively), 
at $b >50^\circ$. The total rms foreground fluctuations (solid curve) are, 
however, dominated, for small angular scales, by extragalactic sources. 
For radiogalaxies we have exploited the baseline model by Toffolatti et 
al. (1998), except that the steepening of the spectral indices to 
$\alpha = 0.75$ has been assumed to occur at $\nu = 1000\,$GHz. The 
dotted lines show the 
contributions of radiogalaxies plus dusty galaxies, adopting, for the latter, 
the baseline model of Toffolatti et al. (1998; lowest curve), model E by 
Guiderdoni et al. (1998) and the model by Blain et al. (1998). The predictions 
of the last two models are very close to each other; those by Guiderdoni et 
al. (1998) are slightly lower except at the highest frequencies. Fluctuations 
were computed assuming that sources brighter than 1 Jy can be identified and 
removed. 
In order to compute the total foreground fluctuations we have adopted the 
model by Blain et al. (1998). 
The filled circles on the solid curve identify the 
central frequencies of Planck channels. The horizontal line shows, 
for comparison, the rms fluctuation amplitude predicted by the standard 
CDM model.  
 }
\leftskip=0truecm
\rightskip=0truecm
\endinsert

The angular power spectrum of Galactic emission, however, falls rapidly at 
small angular scales ($C_\ell \propto \ell^{-3}$: Gautier et al. 1992, Kogut 
1996, Wright 1998; somewhat flatter slopes have been 
reported for some regions of the sky, cf. Lasenby 1996, Schlegel et al. 1998), 
while a Poisson distribution of extragalactic point sources produces a 
white-noise power spectrum, with the same power on all multipoles (Tegmark \& 
Efstathiou 1996). The estimates by Toffolatti et al. (1998) indicate that, at 
high galactic latitude, foreground fluctuations are dominated by extragalactic 
sources for scales $\lsim 30'$. The minimum moves to somewhat higher 
frequencies. Our calculations indicate that, for fluctuations in a beam of 
width $\theta \lsim 30'$ the minimum is close to $100\,$GHz 
while the minimum in the temperature power spectrum for the corresponding 
multipoles ($\ell > 300$) is close to 
$150\,$GHz (see Figs. 1 and 2). The figures also show that at $\ell \simeq 
2000$ the amplitude of the power spectrum of primordial CMB fluctuations is 
expected to be close or below that of foreground fluctuations and the 
situation worsens at larger values of $\ell$ (angular scales smaller than 
$5'$, which are beyond the angular resolution reachable by Planck instruments). 
On the other hand, for smaller values of $\ell$ 
foreground fluctuations are 
not a severe hindrance for measurements of CMB anisotropies for frequencies 
around 100 GHz.


\topinsert
\par\noindent
\centerline{\psfig{figure=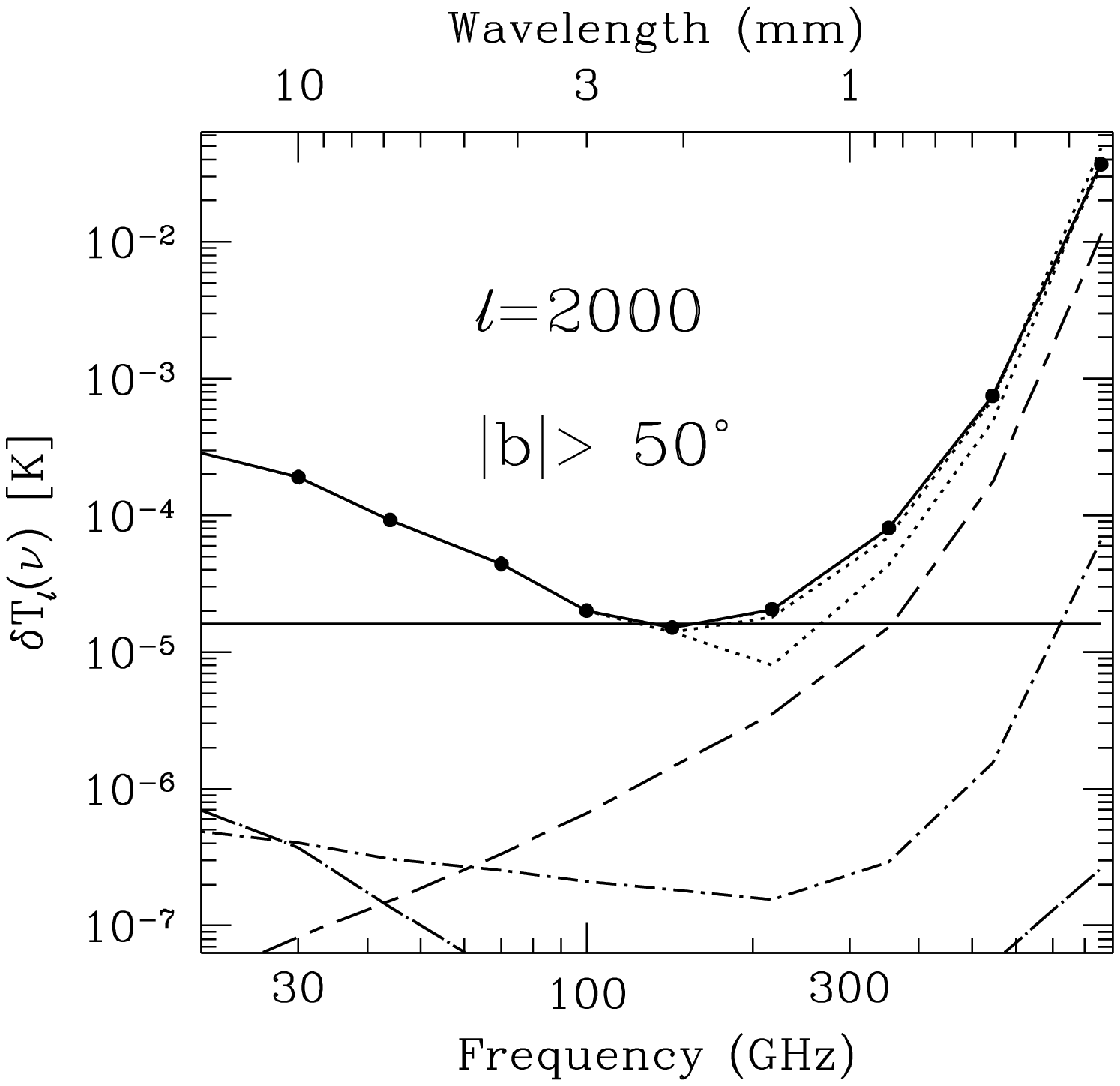,height=9.0truecm,width=12truecm}}


\leftskip=1truecm
\rightskip=1truecm
\vskip-2.truecm
\noindent
{\abstract FIGURE 2. Frequency dependence of CMB brightness fluctuations 
for the multipole $\ell = 2000$, corresponding to an angular scale 
$\theta \simeq 5'$. The lines have the same meaning as in Fig. 1. The dotted 
line based on the model by Blain et al. (1998) is not discernible because it  
lies upon the solid line representing the total foreground fluctuations. 
 }
\leftskip=0truecm
\rightskip=0truecm
\endinsert


%

Fluctuations due to clustering are generally negligible in comparison 
with Poisson fluctuations. However, if discrete sources can be 
subtracted out to flux limits $< 100\,$mJy, the effect of clustering on the 
power spectrum can show up at several tens of arcmin (Toffolatti et al. 
1998).

}

\bsk

{\text                                         
\ni 3. STUDYING EXTRAGALACTIC POINT SOURCES WITH PLANCK

\ni
Just because Planck is designed and optimized to measure CMB 
anisotropies, its sensitivity to compact sources is poor compared 
to ground based instruments, as well as to space-based missions such as FIRST.
Nevertheless, because of the sensitivity near 
fundamental physical limits, freedom from systematic errors, 
full sky coverage, accurate 
calibration from the motion of the Earth around the Sun, 
its images at 9 frequencies, between 30 and 860 GHz 
will be a unique resource for studying both Galactic 
and extragalactic foregrounds over an essentially unexplored, broad 
frequency range.

From several hundreds to many thousands of sources will stand out 
in each Planck channel at more than $5\,\sigma$ (Toffolatti et al. 1998; 
Guiderdoni et al. 1998).
By applying suitable filters, exploiting the point-like nature of 
sources as opposed to the extended CMB and Galactic signals, 
it is possible to suppress by a substantial factor the latter components, 
thus strongly increasing the number of discrete sources that 
can be detected (Tegmark \& de Oliveira-Costa 1998). 
The detectability of discrete sources can be further enhanced by 
exploiting their different spectral properties, in comparison with the 
CMB, i.e. by combining the information from Planck maps at different 
frequencies. 

\smallskip

\ni {\bf 3.1 Evolution of galaxies}

\ni 
Measurements of the far-IR to mm extragalactic background intensity,  
deep ISO counts at $175\,\mu$m, and counts at $850\,\mu$m with SCUBA on JCMT 
indicate extreme evolution of galaxies in this 
region, much stronger than in the optical band. This implies that a large 
fraction (perhaps $> 80\%$: see Hughes et al. 1998) 
of the star formation activity 
in the high redshift Universe may have been missed by optical studies.

This conclusion is further supported by evidences of strong dust obscuration 
of $z> 2$ Lyman break galaxies in the Hubble Deep Field found by analyses  
of their optical-IR spectral energy distribution (Sawicki \& Yee 1998; 
Mobasher \& Mazzei 1998; Meurer et al. 1997). 
It follows that far-IR to mm surveys are essential to provide true 
measurements of the early cosmic star formation history (Guiderdoni et al. 
1997, 1998; Franceschini et al. 1997; Burigana et al. 1997; 
De Zotti et al. 1998; Rowan-Robinson et al. 1997).

As mentioned above, current models give widely different predictions for counts 
in the high frequency Planck channels. Correspondingly large differences 
come out in the predicted redshift distributions of detected sources (see, 
e.g., Fig. 19 of Guiderdoni et al. 1998). Thus, Planck will provide valuable  
data, complementary to the deeper surveys from the ground or from the ESA 
FIRST mission, to discriminate among the various evolutionary scenarios.

\smallskip

\ni {\bf 3.2 Physics of radio sources}

\ni 
Large area radio continuum surveys above 30 GHz are not feasible from the 
ground (the beam area of a given telescope scales as $\nu^{-2}$) and no 
space-borne survey experiment operating at frequencies 
below several hundred GHz is foreseen. Thus,  
the information provided by Planck surveys will be unique.

Planck covers the frequency range where the shape of the spectral energy 
distribution of Active Galactic Nuclei is least known and where 
important spectral features, 
carrying essential information on physical conditions of sources,  
show up. We will briefly mention some examples.

\smallskip
\ni {\it Synchrotron self-absorption frequency in compact regions.} As 
already mentioned, observations at 
mm/sub-mm wavelengths often reveal the transition from optically thick 
to optically thin radio emission in the most compact regions and therefore 
provide quantitative information on their physical conditions.
Planck will, for example, allow 
to investigate if there are systematic differences in the synchrotron 
turnover frequencies between e.g. BL Lacs and quasars, as would be 
expected if BL Lacs are angled closer to our line of sight so that 
their turnovers are boosted to higher frequencies, and if there 
are correlations between turnover frequency and luminosity, which is 
also boosted by relativistic beaming effects.

\smallskip
\ni {\it Early phases of radio flares.}
Major high radio frequency flares have been observed in several 
compact radio sources. The radio emission peak frequency of the quasar 
PKS~0528$+$134 has increased, in 1992, from $\simeq 7\,$GHz to 
$\simeq 60\,$GHz (Zhang et al. 1994). The peak frequency of 
3C~345 has been observed to decrease with time from $\simeq 50\,$GHz
to $\simeq 10\,$GHz. Planck might detect the rise of the flare at the highest 
frequencies, generally missed by ground based observations.

Establishing the peak of the synchrotron emission is crucial also to 
understand if the emission at higher energies is to be attributed 
to Compton scattering of the same synchrotron photons (synchrotron 
self-Compton, SSC) or of seed photons external to the synchrotron 
emitting region, EC). 

\smallskip
\ni {\it Steepening of the synchrotron spectrum due to   
radiation energy losses by the relativistic electrons.}
The spectral break frequency, $\nu_b$, at which the synchrotron 
spectrum steepens, is related to the magnetic field $B$  
and to the ``synchrotron age'' $t_s$ (in Myr) by (e.g. Carilli et al. 1991): 
$\nu_b \simeq 96 (30\,\mu\hbox{G}/B)^{3}t_s^{-2}\,\hbox{GHz}$. 
Thus, the systematic multifrequency 
study at the Planck frequencies will provide a 
statistical estimate of the radiosource ages.

\smallskip
\ni {\it Excesses due to cold dust emission.}  
Excess far-IR/sub-mm emission, possibly due to dust, is often observed from 
radio galaxies (Knapp \& Patten 1991). Planck data will allow to assess 
whether this is a general property of these sources; 
this would have interesting implications 
for the presence of interstellar matter in the host galaxies, 
generally identified with giant ellipticals, which are 
usually thought to be devoid of interstellar matter.

\smallskip
\ni Thus, while lower frequency surveys provide much more detailed information 
relevant to define {\it phenomenological} evolution properties, surveys 
at mm wavelengths are unique to provide information on the {\it physical} 
properties.

\smallskip

\ni {\bf 3.3 Inverted spectrum sources}

\ni 
Planck/LFI will also provide the first complete samples 
of the extremely interesting classes of extragalactic radio sources 
characterized by inverted spectra (i.e. flux density increasing with 
frequency), which are difficult to detect, and therefore  
are either missing from, or strongly 
underepresented in low frequency surveys.
Strongly inverted spectra up to tens of GHz can be produced 
in very compact, high electron density regions, by  
optically thick synchrotron emission or by free-free absorption. 
Examples are known also among galactic sources. 

\smallskip
\ni {\it GHz Peaked Spectrum (GPS) radio sources.} These sources are important 
because they may be the younger stages of radio source evolution 
(Fanti et al. 1995; Readhead et al. 1996) and may thus provide insight 
into the genesis and evolution of radio sources; alternatively, they 
may be sources which are kept very compact by unusual conditions 
(high density and/or turbulence) in the interstellar medium of the 
host galaxy (van Breugel et al. 1984).

\smallskip
\ni {\it High frequency free-free self absorption cutoffs in AGNs.}
Ionized gas in the nuclear region free-free absorbs radio photons 
up to a frequency 
$$\nu_{\rm ff} \simeq 50 (g/5) \left(n_e/10^5\,{\rm cm}^{-3}\right)
\left({{\rm T}/10^4\,{\rm K}}\right)^{-3/4} l_{\rm pc}^{1/2}\ {\rm GHz},$$
where $g$ is the Gaunt factor.
Free-free absorption cutoffs at frequencies $> 10\,$GHz may indeed be 
expected in the framework of the standard torus scenario for unifying 
type 1 and type 2 AGNs, for radio cores seen edge on, and may have been 
observed in some cases (see Barvainis \& Lonsdale 1998).  
Thus Planck's high radio frequency observations can 
provide constraints on physical conditions in the parsec scale accretion disk 
or infall region for the nearest AGNs. 

\smallskip
\ni {\it Advection-dominated sources.}
In the case of a low accretion rate into a massive black hole, the 
radiative cooling rate becomes smaller than the viscous heating rate. 
As a result, the dissipated accretion energy in not efficiently 
radiated away but kept as internal heat and advected inward with 
the accreted plasma (Yi \& Boughn 1998 and references therein). 
The radio spectra of ADS sources are characterized by a synchrotron 
self absorption frequency:
$$\nu_s \simeq 300 \left({M_{\rm BH}/10^8\,{\rm M}_\odot}
\right)^{-1/2}\left[({\dot{M}/\dot{M}_{\rm Edd})/10^{-3}}\right]^{1/2}
\left({T/10^9 K}\right)^2 \left({R/R_S}\right)^{-5/4}\ {\rm GHz}.$$ 
The luminosities, however, are rather low, so that Planck may eventually see 
only the nearest sources of this class. 

}

\bsk

{\text                                         
\ni 4. CONCLUSIONS

\ni
Although current estimates of fluctuations due to extragalactic sources 
are still affected by considerable uncertainties, particularly in the case of 
far-IR/sub-mm sources, conservative estimates allow to conclude that, in the 
frequency range 100--200 GHz, foreground fluctuations over much of the sky 
are well below the expected amplitude of CMB fluctuations on all angular 
scales covered by the Planck mission ($\theta \geq 5'$).

Foreground temperature fluctuations in a Gaussian beam of width $\theta$  
are minimum between 60 and  
$100\,$GHz, higher frequencies corresponding to higher Galactic latitudes 
and smaller values of $\theta$. 
The minimum in the foreground power spectrum is estimated to 
occur at about 150 GHz for $\ell \gsim 300$.

The contribution due to clustering is generally small in comparison 
with the Poisson term; however, the relative importance of clustering 
increases if sources are subtracted from the Planck maps 
down to faint flux levels.

Planck will carry out calibrated all sky surveys at 9 frequencies 
between 30 and 860 GHz, covering an essentially unexplored spectral region.
Sub-mm Planck channels will detect a large number of galaxies up 
to substantial redshifts and will thus provide information on the 
star formation history of galaxies.
Planck will also provide unique information on the physics of 
compact radio sources and in particular: 
on the physical conditions in the most compact components (transition 
from optically thick to optically thin synchrotron emission, ageing of 
relativistic electrons, high frequency flares) and on their relationship 
with emissions at higher energies (SSC versus EC models);
on the frequency of sub-mm excesses due to dust; on  
the population properties of inverted spectrum sources: 
GPS, sources with high frequency free-free self absoprtion, ADS, ...

In conclusion, extragalactic sources will not be a threat to Planck's 
cosmological investigations. At the same time, Planck will provide 
extremely interesting data for astrophysical studies.

\medskip\noindent
{ACKNOWLEDGEMENTS.} We thank the organizing committee for their warm 
hospitality. B. Guiderdoni very kindly sent us the counts predicted by his 
model E in tabular form. 
We gratefully acknowledge the long-standing, very fruitful 
collaboration on extragalactic foregrounds with L. Danese, A. Franceschini, 
P. Mazzei and C. Burigana. LT thanks F. Arg\"ueso G\`omez for helpful 
discussions and suggestions. 
This research has been supported in part by 
grants from ASI and CNR. LT acknowledges partial financial support from the 
Spanish DGES, projects PB95--1132--C02--02 and PB95--0041.

}

  \bsk
  
  \vskip 0.1truecm
  \ni {REFERENCES}
  \ssk
{  

\ref Banday A.J., et al., 1996, {\sl Ap.J.}, {\bf 468}, L85

\ref Barger A.J., et al., 1998, {\sl Nature}, {\bf 394}, 248

\ref Barvainis R. \& Lonsdale C., 1998, {\sl A. J.}, {\bf 115}, 885

\ref Bertin E., Dennefeld M. \& Moshir M., 1997, {\sl A\&A}, {\bf 323}, 685
  
\ref Blain A.W., Ivison R.J. \& Smail I., 1998, {\sl M.N.R.A.S.}, {\bf 296}, 
L29

\ref Burigana C., et al., 1997, {\sl M.N.R.A.S.}, {\bf 287}, L17

\ref Carilli C.L., Perley R.A., Dreher J.W. \& Leahy J.P., 1991, {\sl Ap.J.}, 
{\bf 383}, 554

\ref Chini R., Kr\"ugel E., Lemke R. \& Ward-Thompson D., 1995, {\sl A\&A}, 
{\bf 295}, 317

\ref Condon J.J., 1984, {\sl Ap. J.}, {\bf 287}, 461


\ref Danese L., De Zotti G., Franceschini A. \& Toffolatti L., 1987, {\sl Ap. 
J.}, {\bf 318}, L15

\ref De Zotti G., et al., 1998, A.S.P. Conference Series, {\bf 146}, 158

\ref Eales S., et al., 1998, submitted to Ap. J. Lett., astro-ph/9808040

\ref Fanti C., et al., 1995, {\sl A\&A}, {\bf 302}, 317 

\ref Fixsen D.J., Dwek E., Mather J.C., Bennett C.L. \& Shafer R.A., 1998, 
{\sl Ap. J.}, in press

\ref Franceschini A. \& Andreani P., 1995, {\sl Ap. J.}, {\bf 440}, L5

\ref Franceschini A., Andreani P. \& Danese L., 1998, {\sl M.N.R.A.S.}, 
{\bf 296}, 709

\ref Franceschini A., et al., 1997, ESA SP-401, 159

\ref Franceschini A., Toffolatti L., Danese L. \& De Zotti G., 1989, 
{\sl Ap.J.}, {\bf 344}, 35 

\ref Gautier T.N.I, Boulanger F., Perault M. \& Puget J.L., 1992, {\sl A. J.}, 
{\bf 103}, 1313

\ref Gawiser E. \& Smoot G.F., 1997, {\sl Ap.J.}, {\bf 480} L1 

\ref Gregorich D.T., et al., 1995, {\sl A. J.}, {\bf 110}, 259
 
\ref Guiderdoni B., et al., 1997, {\sl Nature}, {\bf 390}, 257 

\ref Guiderdoni B., Hivon E., Bouchet F.R. \& Maffei B., 1998,
{\sl M.N.R.A.S.}, {\bf 295}, 877

\ref Hacking P. \& Houck J.R., 1987, {\sl Ap. J. Suppl.}, {\bf 63}, 311

\ref Hauser M.G., et al., 1998, {\sl Ap. J.}, in press, astro-ph/9806167

\ref Holdaway M.A., Owen F.N. \& Rupen M.P., 1994, NRAO report

\ref Hughes D.H., et al., 1998, {\sl Nature}, {\bf 394}, 241

\ref Impey C.D. \& Neugebauer G., 1988, {\sl A. J.}, {\bf 95}, 307 

\ref Kawara K., et al., 1997, ESA SP-401, 285

\ref Kellermann K.I., 1964, {\sl Ap. J.}, {\bf 140}, 969 

\ref Kellermann K.I. \& Pauliny-Toth I.I.K., 1971, {\sl Ap. Lett.}, {\bf 8}, 
153 

\ref Knapp G.R. \& Patten B.M., 1991, {\sl A. J.}, {\bf 101}, 1609

\ref Kogut A., 1996, in ``Microwave Background Anisotropies'', eds F.R. 
Bouchet, R. Gispert \& B. Guiderdoni, Ed. Fronti\`eres, p. 445

\ref Kogut A., et al., 1994, {\sl Ap.J.}, {\bf 433}, 435

\ref Kogut A., et al., 1996, {\sl Ap.J.}, {\bf 464}, L5

\ref Lasenby A. N., 1996, in ``Microwave Background Anisotropies'', eds F.R. 
Bouchet, R. Gispert \& B. Guiderdoni, Ed. Fronti\`eres, p. 453

\ref Meurer G.R., et al., 1997, {\sl A. J.}, {\bf 114}, 54   

\ref Mobasher B. \& Mazzei P., 1998, {\sl M.N.R.A.S.}, submitted

\ref Owen F.N. \& Mufson S.L., 1977, {\sl A.J.}, {\bf 82}, 776

\ref Partridge R.B., et al., 1997, {\sl Ap. J.}, {\bf 483}, 38

\ref Puget J.-L., et al., 1996, {\sl A\& A}, {\bf 308}, L5 

\ref Puget J.L., et al., 1998, {\sl A\& A}, submitted

\ref Readhead A.C.S., Taylor G.B., Pearson T.J., Wilkinson P.N., 1996, 
{\sl Ap. J.}, {\bf 460}, 634

\ref Rowan-Robinson M., 1998, preprint, to be submitted to {\sl M.N.R.A.S.}

\ref Rowan-Robinson M., et al., 1997, {\sl M.N.R.A.S.}, {\bf 289}, 490

\ref Rybicki G.B. \& Lightman A.P., 1979, {\sl Radiative Processes in 
Astrophysics}, Wiley

\ref Sawicki M.J. \& Yee H.C.K., 1998, {\sl A. J.}, {\bf 115}, 1329

\ref Schlegel D.J., Finkbeiner D.P., Davis M., 1998, {\sl Ap. J.}, {\bf 500}, 
525

\ref Smail I., Ivison R.J. \& Blain A.W., 1997, {\sl Ap. J.}, {\bf 490}, L5

\ref Stevens J.A., et al., 1996, {\sl Ap. J.}, {\bf 466}, 158 

\ref Tegmark M. \& de Oliveira-Costa A., 1998, {\sl Ap. J.}, {\bf 500}, L83
   
\ref Tegmark M. \& Efstathiou G., 1996, {\sl M.N.R.A.S.}, {\bf 281}, 1297

\ref Toffolatti L., et al., 1998, {\sl M.N.R.A.S.}, {\bf 297}, 117  

\ref van Breugel W., Miley G., Heckman T., 1984, {\sl A. J.}, {\bf 89}, 5

\ref Windhorst R.A., et al., 1993, {\sl Ap. J.}, {\bf 405}, 498

\ref Wright E.L., 1998, {\sl Ap. J.}, {\bf 496}, 1

\ref Yi I. \& Boughn P., 1998, {\sl Ap. J.}, {\bf 499}, 198

\ref Zhang Y.F., 1994, {\sl Ap. J.}, {\bf 432}, 91 

  }                                

\end